\documentclass[twocolumn,showpacs,superscriptaddress,preprintnumbers,amsmath,amssymb,prc]{revtex4-2}
\usepackage{multirow}
\usepackage{amssymb}
\usepackage{amsmath}
\usepackage{graphicx}
\usepackage[normalem]{ulem}
\usepackage{cancel}
\usepackage{multirow}
\usepackage{CJK}
\usepackage[usenames]{color}
\usepackage[colorlinks,linkcolor=blue,urlcolor=blue,citecolor=blue]{hyperref}
\usepackage{tikz,xcolor,hyperref}

\definecolor{lime}{HTML}{A6CE39}
\DeclareRobustCommand{\orcidicon}{
	\begin{tikzpicture}
	\draw[lime, fill=lime] (0,0) 
	circle [radius=0.16] 
	node[white] {{\fontfamily{qag}\selectfont \tiny ID}};
	\draw[white, fill=white] (-0.0625,0.095) 
	circle [radius=0.007];
	\end{tikzpicture}
	\hspace{-2mm}
}
\foreach \x in {A, ..., Z}{%
	\expandafter\xdef\csname orcid\x\endcsname{\noexpand\href{https://orcid.org/\csname orcidauthor\x\endcsname}{\noexpand\orcidicon}}
}

\usepackage{soul,xcolor}
\setstcolor{red}
\newcommand\redsout{\bgroup\markoverwith{\textcolor{red}{\rule[0.5ex]{2pt}{0.4pt}}}\ULon}

\begin{document}
\begin{CJK*} {UTF8} {gbsn}

\title{Isovector giant dipole resonance mode with an improved propagation approach in the framework of EQMD model} 

\author{Chen-Zhong Shi(施晨钟)}
\affiliation{Shanghai Institute of Applied Physics,  Chinese Academy of Sciences, Shanghai 201800, China}
 \affiliation{Key Laboratory of Nuclear Physics and Ion-Beam Application (MOE), Institute of Modern Physics, Fudan University, Shanghai 200433, China}
\author{Xiang-Zhou Cai(蔡翔舟)}%
\affiliation{Shanghai Institute of Applied Physics,  Chinese Academy of Sciences, Shanghai 201800, China}%

\author{De-Qing Fang(方德清)\orcidD{}} 
\affiliation{Key Laboratory of Nuclear Physics and Ion-Beam Application (MOE), Institute of Modern Physics, Fudan University, Shanghai 200433, China}
\affiliation{Shanghai Research Center for Theoretical Nuclear Physics， NSFC and Fudan University, Shanghai }

\author{Yu-Gang Ma(马余刚)\orcidC{}} \thanks{Corresponding author:  mayugang@fudan.edu.cn}
\affiliation{Key Laboratory of Nuclear Physics and Ion-Beam Application (MOE), Institute of Modern Physics, Fudan University, Shanghai 200433, China}
\affiliation{Shanghai Research Center for Theoretical Nuclear Physics， NSFC and Fudan University, Shanghai }


\date{\today}

\begin{abstract}
The Extended Quantum Molecular Dynamics (EQMD) model is one of the few QMD-like transport approaches that can describe the $\alpha$-clustering structure with efficient computational power.
However, compared to most QMD-like models, the choice of equation of state (EOS) for nuclear matter is very limited.   
In this work, a Monte Carlo integral method is employed to calculate the density integration with non-integer exponent.
We demonstrate the superiority of our approach by studying the isovector giant dipole resonance (IVGDR).
This improvement will be beneficial for the EQMD model to study more valuable effects for heavy ion collisions in the near future.
\end{abstract}

\maketitle

\section{Introduction\label{introduction}}

Heavy-ion collisions (HICs) provide a unique opportunity to investigate properties of nuclear matter in a broad temperature and density range by  terrestrial laboratories \cite{PPNP,PPNP2,DengXG2024,Annu,PRL,PRL_NST,PhysRep,SCPMA,NST}. 
In the last 50 years, a large number of experimental data for HICs have been obtained.
The related phenomena and the corresponding physical mechanisms are very complicate and challenging,
so there is an increasing need to develop theoretical models to describe the HICs process from Fermi energy to relativistic energy regimes self-consistently.
In this context, the quantum molecular dynamics (QMD) approach \cite{first_qmd} was first developed by Aichelin \textit{et al}.
In the following decades, many kinds of QMD-like transport approaches have emerged.
For examples, the constraint molecular dynamics approach  (CoMD) \cite{CoMD}, the antisymmetrized molecular dynamics  (AMD) \cite{AMD}, the  fermionic molecular dynamics  (FMD) \cite{FMD}, the ultra-relativistic quantum molecular dynamics (UrQMD) \cite{UrQMD}, and various improved versions of quantum molecular dynamics, see eg.
an improved QMD (ImQMD) \cite{ImQMD}, the isospin quantum molecular dynamics  (IQMD) \cite{IQMD}, the isospin-dependent quantum molecular dynamics at BNU (IQMD-BNU) \cite{IQMD_BNU}, as well as the Lanzhou isospin-dependent quantum molecular dynamics (LQMD) \cite{LQMD} etc. 
Compared with the original QMD model, these models have made different improvements in the mean-field potential, nucleon-nucleon (N-N) collision, the treatments of Pauli blocking, even the description of system in phase space. 
For examples, IQMD treats different charge states of nucleons, deltas and pions explicitly, CoMD introduces a numerical constraint on  phase space to maintain system Fermionic nature.
A more direct way is to describe the system by a Slater determinant, such as AMD and FMD.
In Refs. \cite{ImQMD,IQMD,IQMD_BNU}, 
authors introduced various effects into the effective interactions for different purposes.
Accordingly, some sets of potential parameters have been employed. A recent nice review on transport model comparison of intermediate-energy heavy-ion collisions can be found in Ref.~\cite{PPNP}.

EQMD \cite{EQMD} model as one of a few transport approaches can provide a special opportunity to investigate the $\alpha$-cluster effects, which are  currently an interesting topic in nuclear physics as well as astrophysics \cite{Nat1,Nat2,Nat3},   with efficient computing performance  \cite{hewb1,hewb2,Ma_NT}.
It was first presented by Maruyama {\it et al.} in view of the simulation of heavy-ion reaction at low energy due to its outstanding stability for initial state of nucleus. 
Compared with most QMD-like transport models, EQMD model was improved in following  aspects. 
A phenomenological Pauli potential was adopted to enhance the Fermi properties at low energy.
A friction cooling method was used for giving a stable state, or even a ground state of nucleus.
Besides, the dynamical wave packets and  zero-point kinetic energy have also been considered.
In the last nearly 30 years, the EQMD model was used to study different physics topics at low-intermediate energies, such as collective flow \cite{guocc,shicz2}, giant resonance \cite{hewb1,hewb2,CaoYT2022}, bremsstrahlung \cite{shicz1}, photonuclear reaction \cite{huangbs,HuangBS2021}, liquid-gas phase transition \cite{Wada,guocq,CaoYT2023}, short range correlation \cite{shenl} and so on.
Unfortunately, compared with other QMD-like models, the development of EQMD model is  quite limited.
Until now, except a recent work \cite{Shi2023}, only the stiff parameters can be used in the  EQMD model.   
However, it is inconsistent with the current understanding of the equation of state (EOS) of nuclear matter, i.e. the $K_\infty$ is $230 \pm 40$ MeV. In fact, even in 1988, when the QMD approach was first presented, it also provides sets of parameters with $\gamma=7/6$ leading to soft incompressibilities \cite{first_qmd}. 
In this context, it is time to include the soft EOS within EQMD model for a more wide application in heavy-ion collisions.

In this work, we analyse the reasons that hinder the  development of EQMD model and propose a solution.
The article is arranged as follows:  a brief introduction of EQMD model and its corresponding improvements are given in Sect. \ref{method}. The results and discussion are given in Sect. \ref{result}. And a summary is provided in Sect.~\ref{conclusion}.

\section{Model and method\label{method}}

\subsection{EQMD model\label{EQMD}}

In the framework of the EQMD model \cite{EQMD}, the nucleons, i.e., protons and neutrons, are assumed to be Gaussian wave packets with dynamical widths, instead of keeping a constant width as adopted by most QMD-like transport approaches.
The wave function of total system can be written as follows
\begin{equation}\label{wave}
\begin{aligned}
\Psi & =\prod_i \varphi_i\left(\mathbf{r}_i\right) \\
\varphi_i\left(\mathbf{r}_i\right) & =\left(\frac{v_i+v_i^*}{2 \pi}\right)^{3 / 4} \exp \left[-\frac{v_i}{2}\left(\mathbf{r}_i-\mathbf{R}_i\right)^2+\frac{i}{\hbar} \mathbf{P}_i \cdot \mathbf{r}_i\right]. \\
\end{aligned}
\end{equation}
Here $\mathbf{R}_i$ and $\mathbf{P}_i$ are centers of position and momentum belong to the $i$-th nucleon in phase space, $v_i=\frac{1}{\lambda_i}+i\delta_i$ is the complex width of wave packet, respectively.
The corresponding density distribution, or namely the probability distribution, can be expressed as
\begin{equation}\label{eq:density}
\rho_i \left( \mathbf{r} \right) = \frac{1}{\left( \pi \lambda_i \right)^{3/2}} \exp \left[ -\left( \mathbf{r}-\mathbf{R}_i\right)^2/ \lambda_i \right].
\end{equation}
Obviously, the real part $\lambda_i$ can not be $\le$ 0.

Following the time-dependent variation principle \cite{FMD} (TDVP), the propagation of nucleons in  mean-field can be solved by 8$A$-dimension Newton equations as follows
\begin{equation}\label{eq:eqofmotion}
\begin{aligned}
\dot{\mathbf{R}}_i &=\frac{\partial H}{\partial \mathbf{P}_i}+\mu_{\mathrm{R}} \frac{\partial H}{\partial \mathbf{R}_i}, ~\dot{\mathbf{P}}_i=-\frac{\partial H}{\partial \mathbf{R}_i}+\mu_{\mathrm{P}} \frac{\partial H}{\partial \mathbf{P}_i}, \\
\frac{3 \hbar}{4} \dot{\lambda}_i &=-\frac{\partial H}{\partial \delta_i}+\mu_\lambda \frac{\partial H}{\partial \lambda_i}, ~\frac{3 \hbar}{4} \dot{\delta}_i=\frac{\partial H}{\partial \lambda_i}+\mu_\delta \frac{\partial H}{\partial \delta_i}.
\end{aligned}
\end{equation}
Here $\mu_\mathbf{R}$, $\mu_\mathbf{P}$, $\mu_{\lambda}$ and $\mu_{\delta}$ are the friction coefficients with negative values intended to prepare energy minimum states as the initial ground state nuclei in the friction cooling process. 
On the other hand, they maintain strictly at a zero value to ensure energy conservation during heavy-ion collisions. 
$H$ is the system Hamiltonian, which is expressed as follows
\begin{equation}
\begin{aligned}
H & =\left\langle\Psi\left|\sum_i-\frac{\hbar^2}{2 m} \nabla_i^2-\hat{T}_{\text {zero }}+\hat{H}_{\text {Int.}}\right| \Psi\right\rangle \\
& =\sum_i\left[\frac{\mathbf{P}_i^2}{2 m}+\frac{3 \hbar^2\left(1+\lambda_i^2 \delta_i^2\right)}{4 m \lambda_i}-T_{\text {zero }}\right]+H_{\mathrm{Int.}} .
\end{aligned}
\end{equation}
The sum of the first three terms in bracket is the total kinetic energy contributed by the $i$-th particle, and the $H_\text{int.}$ is the corresponding potential experienced in the mean-field. 
For the effective interaction adopted in the EQMD model, it can be written as follows 
\begin{equation}
H_{\text {Int. }}=H_{\text{Sky. }}+H_{\text{Coul. }}+H_{\text{Symm. }}+H_{\text {Pauli}},
\end{equation}
where $H_{\text{Sky.}}$, $H_{\text{Coul.}}$, $H_{\text{Symm.}}$ and $H_{\text{Pauli}}$ represent the Skyrme, Coulomb, Symmetry and Pauli potential, respectively. 
In the original EQMD model, the Skyrme interaction only comprises two-body and three-body or namely density-dependent terms as follows: 
\begin{equation}\label{eq:original_eqmd_H_sky}
H_{\text{Sky. }}=\frac{\alpha}{2 \rho_0} \int \rho^2(\mathbf{r}) \mathrm{d} \mathbf{r}+\frac{\beta}{(\gamma+1) \rho_0^\gamma} \int \rho^{\gamma+1}(\mathbf{r}) \mathrm{d} \mathbf{r},
\end{equation}
where $\alpha$ is related to the two-body term, $\beta$ and $\gamma$ are related to the three-body term.
Specifically, the $\gamma$ is strongly correlated with the incompressibility of the nuclear matter equation of state (EOS).
The symmetry potential utilized as the simplest formula, as follows:
\begin{equation}
H_{\text{Symm.}}=\frac{c_{\mathrm{s}}}{2\rho_0}\int{ \left(\rho_\text{p}-\rho_\text{n}\right)^2 \mathrm{d} \mathbf{r}},
\end{equation}
where $\rho_\text{p}$ and $\rho_\text{n}$ denote the proton and neutron density, respectively.
For the Coulomb interaction, only the direct term is considered
\begin{equation}
H_{\text {Coul. }}=\frac{a}{2} \int \mathrm{d} \mathbf{r} \mathrm{d} \mathbf{r}^{\prime} \frac{\rho_{\mathrm{p}}(\mathbf{r}) \rho_{\mathrm{p}}\left(\mathbf{r}^{\prime}\right)}{\left|\mathbf{r}-\mathbf{r}^{\prime}\right|},
\end{equation}
where $a$ is the fine structure constant.
The Pauli potential~\cite{EQMD} acts as a part of the Fermi properties in an antisymmetrized many-body system.
It can been written as 
\begin{equation} \label{eq:pauli}
H_\text{Pauli}  = \frac{c_P}{2} \sum_i\left(f_i-f_0\right)^\mu \theta\left(f_i-f_0\right), 
\end{equation}
where $c_p$ and $\mu$ are the strength and power of the Pauli potential, $\theta$ is a unit step function,  $f_i$ is the overlap of a nucleon with the identical particle including itself, i.e. $f_i  \equiv \sum_j \delta\left(S_i, S_j\right) \delta\left(T_i, T_j\right)\left|\left\langle\phi_i \mid \phi_j\right\rangle\right|^2$, and $f_0$ is the threshold parameter, which takes a value close to 1.
The $T_\text{zero}$ is the zero-point kinetic energy, and the subtraction of $T_\text{zero}$ intents to avoid a spurious kinetic energies arising from the the zero-point oscillation of the center of mass of isolated fragments or nucleons \cite{AMD}.
Further details about Pauli potential and zero-point kinetic energy can be found elsewhere \cite{EQMD,AMD}.
In original EQMD model, there is no conventional momentum-dependent interaction taken into account.
In a recent study \cite{dixp}, a logarithm type of momentum dependent interaction \cite{momentum_interaction_log_1, momentum_interaction_log_2} was also considered
\begin{equation}
U_{\text {mdi }}=\delta \cdot \ln ^2\left(\varepsilon \cdot(\Delta \mathbf{p})^2+1\right) \cdot\left(\frac{\rho_{\text {int }}}{\rho_0}\right),
\end{equation}
 where $\Delta\mathbf{p}=\mathbf{p}_1-\mathbf{p}_2$ is given in units of MeV c$^{-1}$, $\delta=1.57$ MeV and $\varepsilon=5\times 10^{-4}$ MeV$^{-2}$ c$^2$.

In Tab.~ \ref{tab:table_EQMD}, it displays two sets of original parameters provided by Ref.~\cite{EQMD} and a recent modified configuration from Ref.~ \cite{dixp}.
It is easy to observe an interesting fact, no matter which set of parameters employed in the original EQMD model, the variable $\gamma$ can only take an integer value 2 strictly. 
It usually means a very stiff incompressibility of the nuclear matter EOS.
Nowadays, a large number of studies reveals that the EOS favours a softer incompressibility about 230$\pm$40 MeV.
It  indicates $\gamma$ should take a non-integer number less than 2 but great than 1.
Unfortunately, EQMD model has long suffered from a lack of values for which $\gamma$ is a non-integer number.
Therefore it is a crucial question for the development of the EQMD model and we are curious if there are any special reasons preventing $\gamma$ from taking a non-integer value.
In addition, the form of effective interaction adopted in the EQMD model is to some extent somewhat simpler  compared to other QMD-like models, such as ImQMD, LQMD and QMD-BNV etc., from a contemporary point of view.

\begin{table}[bthp]
\caption{\label{tab:table_EQMD}}
Two sets of potential parameters in original EQMD model  \cite{EQMD} (EQMD 1 and EQMD 2) and a modification by Ref.~\cite{dixp} (EQMD Di).
\begin{ruledtabular}
\begin{tabular}{lrrrr}
\textrm{}&
\textrm{EQMD 1  }&
\textrm{EQMD 2  }&
\textrm{EQMD Di }\\
\colrule
$\alpha$ (MeV)        &  -116.6 &  -124.3&   -134.6\\
$\beta$ (MeV)         &  70.8   &    70.5&     62.6\\
$\gamma$              &  2.0    &     2.0&      2.0\\
$c_\text{s}$ (MeV)    &  25.0   &    25.0&     28.0\\
$c_\text{Pau.}$ (MeV)&  15.0   &    15.0&     28.0\\
$f_0$                 &   1.05  &     1.0&      1.0\\
$\mu$                 &  2.0    &     1.3&      1.3\\
\end{tabular}
\end{ruledtabular}
\end{table}

\begin{table*}[bthp]
\caption{\label{tab:table_skyrme}}
Some common parameter settings and corresponding physical variables \cite{parameters_setting_feng,parameters_setting_zhang}.
\begin{ruledtabular}
\begin{tabular}{lrrrrrr}
\textrm{Force}&
\textrm{SkP }&
\textrm{SkM* }&
\textrm{RATP }&
\textrm{SLy4 }&
\textrm{SLy6 }&
\textrm{SLy7 }\\
\colrule
$\alpha$ (MeV)             & -356.2 &-317.4 &-259.2 &-298.7 &-295.7 &-294.0  \\ 
$\beta$ (MeV)              &  303.0 & 249.0 & 176.9 & 220.0 & 216.7 & 215.0  \\
$\gamma$                   &  7/6   & 7/6   & 1.2   & 7/6   & 7/6   & 7/6    \\
$g_\text{Sur.}$ (MeV)      &  19.5  & 21.8  & 25.6  & 24.6  & 22.9  & 22.6   \\
$g_\text{Sur.}^{iso}$ (MeV)&  -11.3 & -5.5  & 0.0   & -5.0  & -2.7  &-2.3    \\
$g_{\tau}$ (MeV)           &  0.0   & 5.9   & 11.0  & 9.7   & 9.9   & 9.9    \\
$c_s$ (MeV)                &  30.9  & 30.1  & 29.3  &32.0   &32.0   & 32.6   \\
$\rho_{\infty}$ (fm$^{-3}$)&  0.162 &0.160  &0.160  &0.160  &0.160  &0.158   \\
$m^{\prime}/m$             &  1.00  & 0.79  &0.67   &0.70   &0.69   &0.69    \\
$K_\infty$ (MeV)           &  200   &216    &239    &230    & 230   &229     \\
\end{tabular}
\end{ruledtabular}
\end{table*}

\subsection{Modification of mean-field potential \label{improve}}

In the last 30 years, the QMD-like transport approaches have undergone significant developments.
There are many authors and groups have made great contributions on QMD-like model.
A typical template of the effective interaction, widely used in the past decades, can be written as follows \cite{Feng_2011,zhang_2012}
\begin{equation}\label{eq:new}
\begin{aligned}
u_\text{Int.}(\rho)&=  \frac{\alpha}{2} \frac{\rho^2}{\rho_0}+\frac{\beta}{(\gamma+1) \rho_0^\gamma} \rho^{\gamma+1}\\
&+\frac{g_{\text {Surf.}}}{2 \rho_0}(\nabla \rho)^2+\frac{g_{\text {Surf.}}^{\text {iso }}}{2 \rho_0}\left[\nabla\left(\rho_{\mathrm{n}}-\rho_{\mathrm{p}}\right)\right]^2 \\
& +\frac{c_s}{2 \rho_0^{\gamma_s}} \rho^{\gamma_s-1} \left(\rho_\text{n}-\rho_\text{p}\right)^2+g_\tau \frac{\rho^{8 / 3}}{\rho_0^{5 / 3}}.
\end{aligned}
\end{equation} 
Here $\alpha$, $\beta$ and $\gamma$ are the same as in Eq. \ref{eq:original_eqmd_H_sky}.
The $g_\text{Surf.}$ and $g_\text{Surf.}^\text{iso}$ are the coefficients related to the surface interaction and corresponding symmetry part, and $g_{\tau}$ is related to the momentum-dependent interaction.
In the developmental history of QMD, there is a diversity of mathematical forms concerning the momentum dependent interactions. 
Here, a density dependent form, derived from the local density  approximation, is utilized. 
The $c_\text{s}$ and $\gamma_\text{s}$ are related to the symmetry energy interaction.
Generally speaking, the EOS is labeled as stiff-asy for $\gamma_\text{s}>1$ and as soft-asy for $\gamma_\text{s}<1$.
In Ref. \cite{double_ratio_np_buu,gamma_05}, they found that a soft-asy is suitable for  potential through the neutron-proton spectral double ratios. 
Base on this reason, we take $\gamma_\text{s}=0.5$ in the rest of article unless otherwise specified.
Moreover, for the Coulomb interaction, an exchange part is also considered \cite{coul_exchange},
\begin{equation}
u_{\text{Coul.}}(r) = \frac{a}{2} \int \mathrm{d}^3 r^{\prime} \frac{\rho_{\mathrm{p}}(r)\rho_{\mathrm{p}}\left(r^{\prime}\right)}{\left|r-r^{\prime}\right|}-\frac{3}{4} e^2\left(\frac{3}{\pi}\right)^{1 / 3}  \rho_{\mathrm{p}}^{4 / 3}(r).
\end{equation}

In Tab.~\ref{tab:table_skyrme}, it displays 6 widely utilized sets of parameters and some corresponding physical variables.
 $\rho_0$, $m^{\prime}$ and $K_\infty$ represent saturated density of infinity nuclear matter, effective mass and the incompressibility, respectively.
$K_\infty$ is related to $\beta$, $\gamma$, $m^\prime$ and $\rho_0$.
Its definition can be written in the following form
\begin{equation}
K_{\infty}=\left.9 \rho_0 \frac{\partial^2(E / A)}{\partial \rho^2}\right|_{\rho=\rho_0} .
\end{equation}
Here $E/A$ is the binding energy per nucleon of infinity nuclear matter, and the definition can be found in elsewhere.
Then the $K_\infty$ can be directly deduced as follow
\begin{equation}
K_{\infty}=k \rho_0^{2 / 3}\left(10 \frac{m}{m^{\prime}}-12\right)+\frac{9 \beta \gamma(\gamma-1)}{\gamma+1},
\end{equation}
where $k\approx 75 \mathrm{MeV fm^2}$. 
Using the comparison of EQMD set 2 and SkP as an example, the other situation are similar.
It is not difficult to see that although $\beta$ has increased from 70.5 MeV to 303.0 MeV, with the change in $\gamma$ from 2 to 7/6, the overall value of incompressibility is still decreasing.
However, it inevitably encounters the problem that the nucleon propagation has no analytical solution when $\gamma$ takes a non-integer value between 1 and 2.
For many QMD-like models, such as ImQMD, LQMD and IQMD-BNU, an approximate method is used to solve this problem, as follows 
\begin{equation}
\begin{aligned}
H_\gamma&=\frac{\beta}{(\gamma+1) \rho_0^\gamma}\left\langle\rho^\gamma\right\rangle
\approx \frac{\beta}{(\gamma+1) \rho_0^\gamma}\sum_i\langle\rho\rangle^\gamma_i \\
\langle\rho\rangle^{\gamma}_i&=\sum_{j \neq i} \int{\rho_i \rho_j \mathrm{~d}^3 r}.
\end{aligned}
\end{equation}
This approximate treatment rarely seems to cause significant negative effects in most of the QMD-like approaches mentioned above from previous literature.
However, in the context of EQMD it causes a serious numerical problem leading to initialisation failure or system instability.
One can refer to the literature for relevant reports \cite{dixp,wangning}.
One reason for this is that this approximation usually causes the real parts of the wavenumbers to become negative values as the nucleons propagate. 
For simplicity, we will call this treatment Method I.

In order to solve this question in the framework of EQMD, a Monte Carlo integral method is adopted for the three-body interaction.
The corresponding force felt by the $i$-th particle can be obtained through the partial derivation with respect to $\mathbf{R}_i$ and $\lambda_i$, as follows
\begin{equation} \label{eq:method}
\begin{aligned}
\frac{\partial H_\gamma}{\partial \mathbf{R}_i}=\frac{\beta}{\left(\gamma+1\right)\rho^{\gamma}_0}\int \frac{\partial}{\partial\mathbf{R}_i}\rho^{\gamma+1}\mathrm{~d}\mathbf{r}, \\
\frac{\partial H_\gamma}{\partial \lambda_i}=\frac{\beta}{\left(\gamma+1\right)\rho^{\gamma}_0}\int \frac{\partial}{\partial \lambda_i}\rho^{\gamma+1}\mathrm{~d}\mathbf{r}.
\end{aligned}
\end{equation}
Here the partial differential sign is incorporated into the integral formula. 
Since the $\rho_i$ is a Gaussian form, it is easy to implement a Monte Carlo integral method for this integration. 
We call this treatment as Method II.
A similar method for the accurate calculation of the three-body interaction is first reported in Ref.~\cite{Yang}. 
In the actual numerical calculation of the real system, we also exclude the self-interaction as in Ref.~\cite{EQMD}.
For the same reason, the symmetry term and the momentum dependence term are also a potential source of the numerical problem.
The corresponding treatment is similar to the three-body term described in Eq. ~\ref{eq:method}.

\subsection{Isovector giant dipole resonance\label{gdr}} 

In a classical picture, isovector giant dipole resonance can be qualitatively understood as a collective vibration between centroids of protons and neutrons  \cite{first,Garg,Paar,taochen,hewb1,WangSS2017,Colo_NST,WangSS2023,SongYD2023}. 
Usually this dipole resonance mode was excited by heavy-ion beams \cite{hic}, but also possible with  available $\gamma$-beam \cite{gamma,Colo_NST}, such as Hi$\gamma$S \cite{Higs} and SLEGS \cite{SLEGS1,SLEGS2,SLEGS3}. 
The corresponding perturbation operator can be written as follows \cite{wangrui},
\begin{equation}
\begin{aligned}
&\hat{H}_\text{Exc.}(t) = \lambda \hat{Q} \delta(t-t^{\prime}) \\
&\hat{Q} = \sum^A_i\hat{q}=\frac{N}{A}\sum_{i}^{Z}\hat{r}_\mathbf{n}-\frac{Z}{A}\sum_{i}^{N}\hat{r}_\mathbf{n},
\end{aligned}
\end{equation}
where $\lambda=25$ MeV/c is a perturbation parameter, $\hat{r}_\mathbf{n}$ is a position operator along a specific direction, denoted as $\vec{n}$. 
For the sake of simplicity, only the pseudo-spherical structure nucleus is considered in this work.
So we set the vector $\vec{n}$ along the $z$-axis without loss of generality. 
In order to excite a ground state nucleus into dipole mode, we utilize a treatment suggested by Urban \cite{Urban}.
Since $\langle\hat{Q}\rangle$ only explicitly contains $\mathbf{R}_i$, the new coordinate of center of wave packet belong to the $i$-th nucleon after a perturbation can be written as follows
\begin{equation}
\mathbf{P}_i^\prime \rightarrow \mathbf{P}_i-\lambda \frac{\partial \langle \hat{Q}\rangle \left(\mathbf{R}_1, \mathbf{R}_2, \ldots, \mathbf{R}_i,\ldots\right)}{\partial \mathbf{R}_i}.
\end{equation}
Base on the linear response theory, the response of excitation of $\Delta\hat{Q}$ after a perturbation on the ground state as a function of time can be written as follows
\begin{equation}
\begin{aligned}
\Delta\langle\hat{Q}\rangle(t) & =\left\langle 0^{\prime}|\hat{Q}| 0^{\prime}\right\rangle(t)-\langle 0|\hat{Q}| 0\rangle(t) \\
& =-\frac{2 \lambda \theta(t)}{\hbar} \sum_F|\langle F|\hat{Q}| 0\rangle|^2 \sin \frac{\left(E_F-E_0\right) t}{\hbar}.
\end{aligned}
\end{equation}
Here $|0\rangle$ and $|0^{\prime}\rangle$ represent the states of nucleus before and after perturbation, $|F\rangle$ and $E_F$ are the energy eigenstate of the excited nucleus and eigen-energy, respectively.
The corresponding strength function $S(E)$ can be defined as a Fourier integral of $\Delta\langle\hat{Q}(t)\rangle$ as follows
\begin{equation}\label{eq:se}
S(E) = -\frac{1}{\pi \lambda} \int_0^{\infty} \text{d} t \Delta\langle\hat{Q}(t)\rangle \sin \frac{E t}{\hbar}.
\end{equation}
 
Another important mechanism for giant resonance within the QMD-like approaches is the binary collisions. 
Recent studies indicate that the N-N collision is a main reason for the GR damping, but dose not influence the peak energy position remarkably \cite{wangrui,xujun,SongYD2023}.
In this work, we focus on the enhanced numerical treatment of three-body term, thus the binary collision is omitted to avoid unnecessary complexity.
 
\begin{figure}[htbp]
\resizebox{8.6cm}{!}{\includegraphics{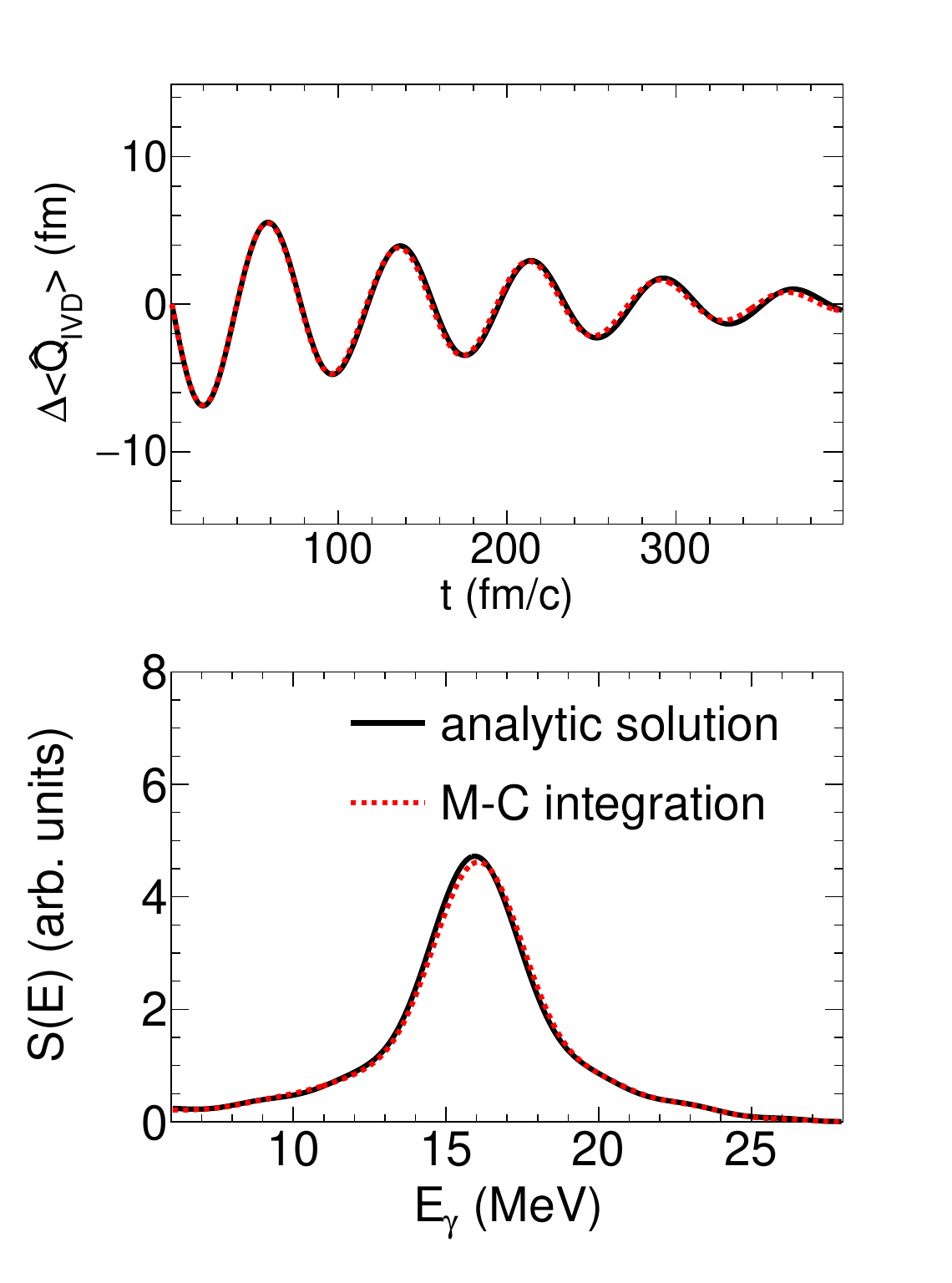}}
\caption{\label{fig:gamma2} Time evolution of $\Delta\langle\hat{Q}\rangle$ (upper panel) and the corresponding strength function (lower panel) for $^{90}$Zr obtained by the analytical solution (solid line) and Monte-Carlo integration method (dashed line).}
\end{figure}

\section{Results and Discussion\label{result}}

\begin{figure}[htbp]
\resizebox{8.6cm}{!}{\includegraphics{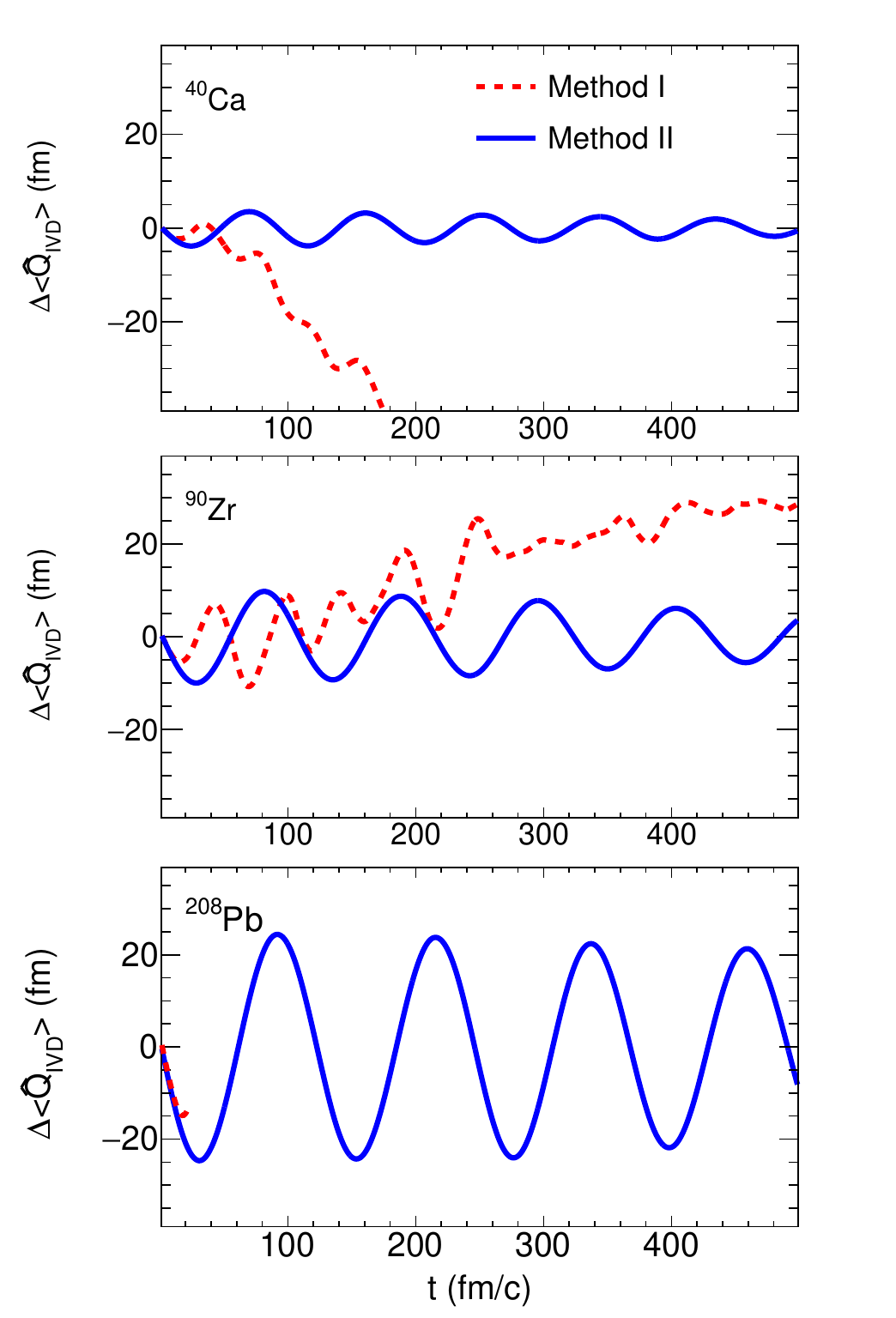}}
\caption{\label{fig:app vs pre} Time evolution of $\Delta\langle\hat{Q}\rangle$ of  $^{40}$Ca, $^{90}$Zr and $^{208}$Pb with two difference treatments for the three-body interaction (Dashed line: Method I, solid line: Method II.)}
\end{figure}

\begin{figure}[htbp]
\resizebox{8.6cm}{!}{\includegraphics{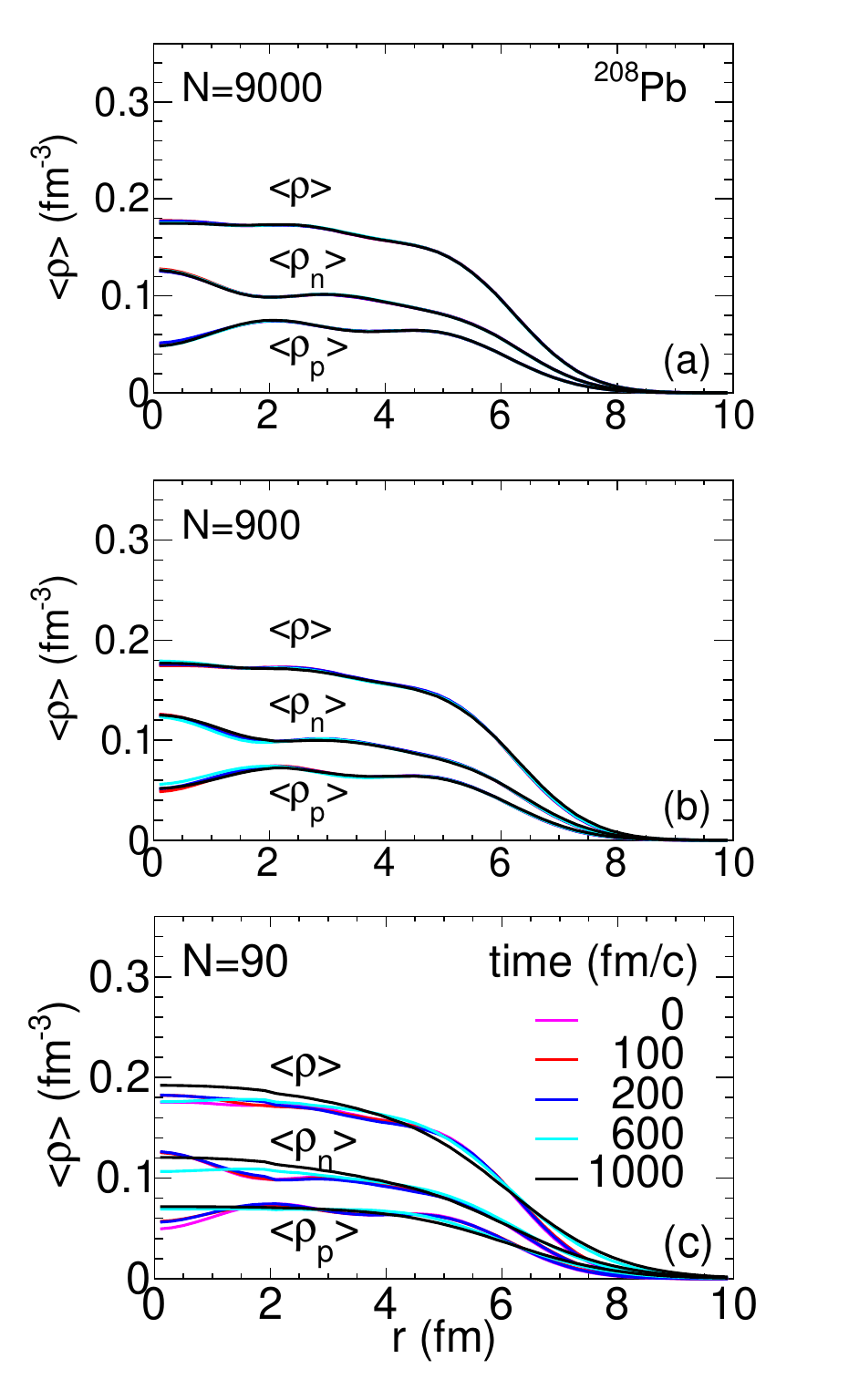}}
\caption{Time evolution of radial density distribution of $^{208}$Pb  during a thousand of fm/c with random number $N$ = 9000, 900, and 90.\label{fig:test number} }
\end{figure}

In the previous works~\cite{hewb1,WangSS2017,HuangBS2021}, the IVGDR can been reproduced reasonably in the framework of EQMD model.
Taking this as a stand point, we extend the application scenario of the EQMD model from $\gamma=2$ to the case of $\gamma$ taking a non-integer number between 1 and 2.
First, the correctness of our numerical method is checked by comparing with the original calculation in EQMD model.
In Fig.~\ref{fig:gamma2}, it displays the time evolution of $\Delta\langle\hat{Q}\rangle$ for $^{90}$Zr.
The solid line is calculated by the original EQMD model with an analytical solution of the three-body interaction.
And the dashed line is obtained by the Method II, that is, the Monte-Carlo integration.
The parameters used in Fig.~\ref{fig:gamma2} is the second EQMD setting as listed in Tab.~\ref{tab:table_EQMD}.
In principle, the dipole oscillation should be the same in those two cases.
In fact, they do coincide well during 0-300 fm/c of the time evolution, and there is a slight deviation in the later evolution.
Consequently, there is a minor variance in the corresponding strength functions. 
Based on the above facts, this means that the Method II is equivalent to the analytical solution when the variable $\gamma$ is 2.

We then extend our model to the case where $\gamma$ is not an integer.
In this situation, the original analytical solution for the three-body interaction is not applicable.
So it is necessary to use some methods to solve this challenge. 
We illustrate a comparison of  Method I and Method II introduced above and the advantage of our numerical solution of three-term integral in Fig.~\ref{fig:app vs pre}.
It displays the time evolution of $\Delta\langle\hat{Q}\rangle$ for the $^{40}$Ca, $^{90}$Zr and $^{208}$Pb.
The dashed line is related to the approximate calculation, i.e. method I; the solid line is related to the Monte Carlo integral method, i.e. method II.
The SkP parameter setting listed in Tab.~\ref{tab:table_skyrme} with $\gamma=7/6$ is adopted for comparison.
Here $\gamma_\text{s}$ is taken as 1.0 is to eliminate the  interference of density integral from the symmetry energy term.
The equations of motion cannot be solved accurately when $\gamma_\text{s}$ taken as a non-integer number.
To simplify the discussion, we fix $\gamma_\text{s}$=1 in this comparison, so that the momentum dependent term and the symmetry energy term can be solved analytically in this configuration.
Any problems occurring during the GDR oscillation can only arise from the density integration associated with the three-body interaction term.
In our opinion the $\gamma_\text{s}$ set to 1.0 can make the conclusions clearer. 

In both cases, the same perturbation, i.e., $\lambda=25$ MeV/c is applied on the same ground state of nucleus. 
It is obviously that our numerical integral method can provide much clear harmonic vibration signals, and this dipole mode can be sustained over a long time.
In contrast, it can not yield a clear harmonic signal and even the diverge occurs rapidly when it is calculated with method I.  
Even in the case of $^{208}$Pb, the simulation encounters a computation error and stops  before the set time.
That is, the dashed line only maintains a very short time in the lower panel. 
This is usually caused by the $\lambda$ associated with one or more nucleons becoming a non-positive number during propagation.
Based on the results above, in the rest of the article we use only method II to treat the propagation of nucleons. 
It should be noted that in both cases of Fig.~ \ref{fig:app vs pre} the same initial nucleus is used as the ground state, initialised by the friction cooling process in method II.
The reason why we refrain from using method I to obtain the ground nucleus is due to the appearance of a serious numerical problem, i.e. the real part of the complex width $\lambda$ becomes negative during the friction cooling process, which will lead to the failure of the initialisation.  

\begin{figure}[htbp]
\resizebox{8.6cm}{!}{\includegraphics{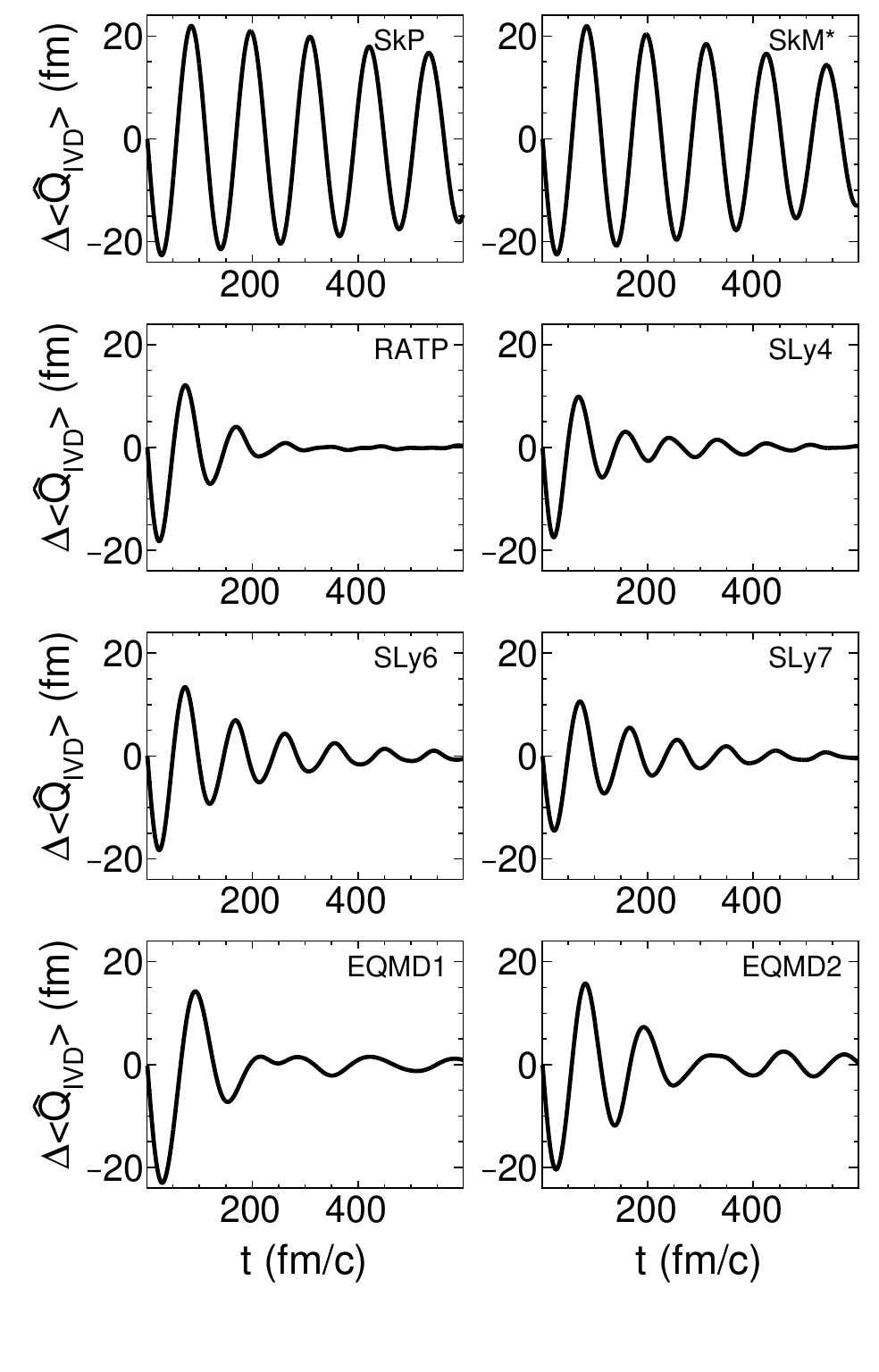}}
\caption{\label{fig:Pb208 vs Skyrme}
The time evolution of dipole vibration with different sets of Skyrme parameters.}
\end{figure}

\begin{figure}[htbp]
\resizebox{8.6cm}{!}{\includegraphics{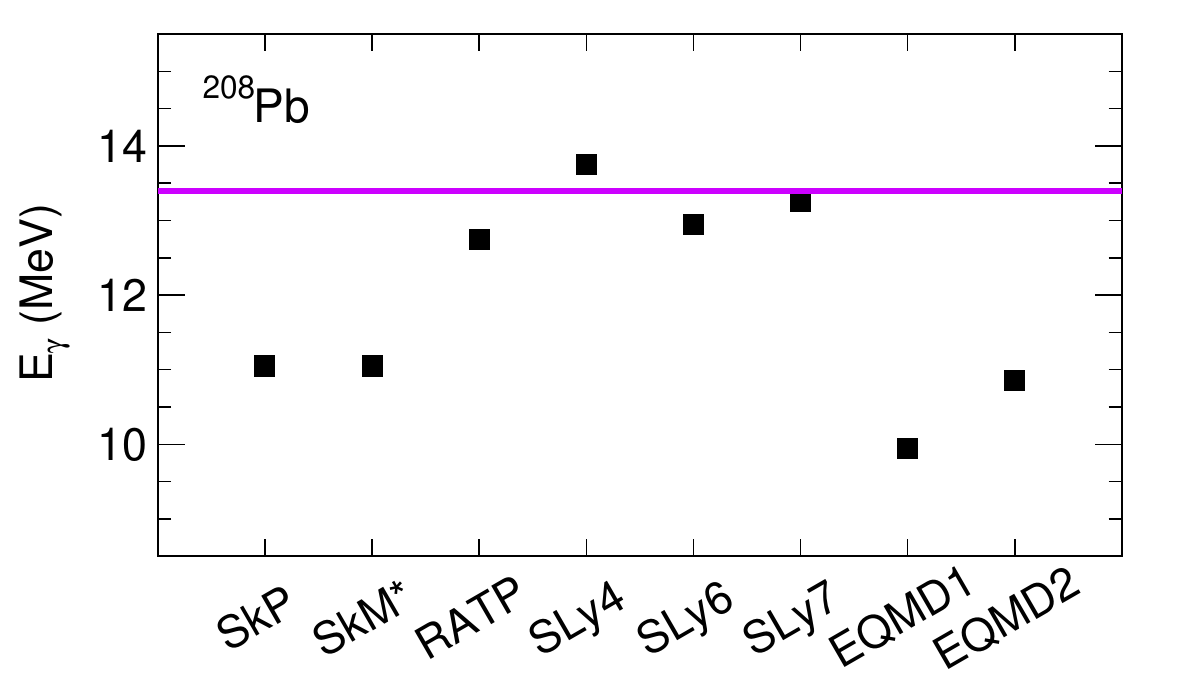}}
\caption{\label{fig:strength of pb} 
The corresponding peak energy with SkP, SkM*, RATP, SLy4, SLy6, SLy7, EQMD 1 and EQMD 2 parameter settings.
The experimental resonance peak is plotted as a horizontal violet line.
}
\end{figure}

It demonstrates a significant advantage of our numerical method for dipole mode calculation in Fig.~ \ref{fig:app vs pre}. 
As is well known, in the context of Monte Carlo integrals, the sample size directly associates with the numerical  accuracy. 
It is therefore crucial to consider the trade-off between the computational overhead and the benefits gained by improving the accuracy.
In Fig. ~\ref{fig:test number}, the time evolution of the radial density in $^{208}$Pb with different number of samples during a thousand of fm/c time span.
As expected, the best stability is observed in the case of 9000 samples, the next one is 900 samples, while the worst one is evident in the 90 samples case.
Based on the results, 900 samples is chosen for a good balance between computational performance and numerical accuracy.

In many previous studies, the GDR of $^{208}$Pb is used to constrain the nuclear properties, such as, nuclear symmetry energy \cite{xujun,songyd}, in-medium N-N cross section \cite{wangrui}, the nucleon effective mass \cite{kong} etc.
So we also choose $^{208}$Pb as a target nucleus to check the 6 different sets of Skyrme parameters in Tab.~\ref{tab:table_skyrme} and the 2 conventional cases of original EQMD in Tab.~\ref{tab:table_EQMD}.
However, one should note that this work does not insist on determining an optimal potential setting for the nuclear matter EOS; instead, we focus on verifying the effectiveness and necessity of our improved algorithm on density integration with non-integer value.
The GDR oscillations with 8 different Skyrme parameter settings are plotted in Fig.~ \ref{fig:Pb208 vs Skyrme}.
Qualitatively  speaking, in the case of SkP and SkM* settings, the vibration amplitudes are larger than other situations, while the damping is smaller than that in the alternative scenarios.
The corresponding peak energies are plotted in Fig.~\ref{fig:strength of pb}. 
The horizontal line represents the resonance peak of 13.4 MeV obtained from $^{208}\text{Pb}\left(p, p^{\prime}\right)$ reaction performed at RCNP \cite{exp_Pb208}.
Obviously, the peak energies of SkP, SkM*, EQMD set 1 and EQMD set 2 are grossly underestimated in comparison with the experimental value.
In fact, in our earlier research, it was often necessary to adopt a larger symmetry energy coefficient to realize reasonable reproductions of GDR for heavy nuclei.
The symmetry energy dependence of the GDR is not the subject of this article and will be the subject of future work. 
For the other EOSs, the peak energies are all close to the experimental value.

\begin{figure}[htbp]
\resizebox{8.6cm}{!}{\includegraphics{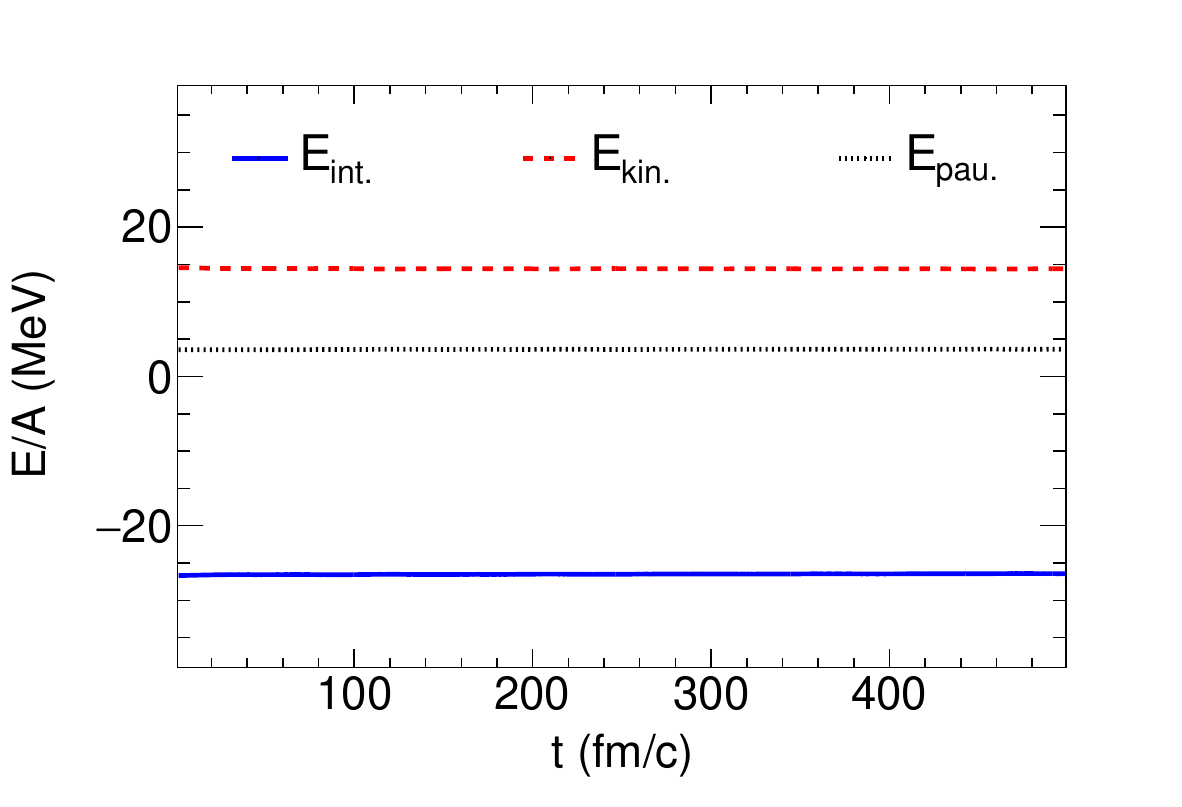}}
\caption{\label{fig:Pauli} The total interaction energy exclusive the Pauli potential, the kinetic energy and the Pauli potential as a function of time during the GDR oscillation of $^{208}$Pb with SLy6 setting.}
\end{figure} 

The fermion properties of nucleons is very important for nuclear structure and low-energy nuclear reaction. 
However, up to now, only the AMD and FMD model treat the Pauli principle strictly with antisymmetrization of phase space which will seriously affect the computational efficiency.
For a good performing, a phenomenological repulsive potential named Pauli potential forbidding the nearby identical particle to come close to each other in the phase space is adopted \cite{EQMD}.
In order to evaluate the effect of Pauli Potential during the GDR oscillation, we plot the total interaction energy per nucleon exclusive Pauil potential, the kinetic energy and the individual Pauli potential as a function of time for the GDR oscillation of $^{208}$Pb with SLy6 setting in Fig.~\ref{fig:Pauli}. 
The absolute value of the Pauli potential is about 4 MeV/u, which is obviously smaller than the total interaction energy of about 26 MeV/u and the kinetic energy of about 15 MeV/u. 

\begin{figure}[htbp]
\resizebox{8.6cm}{!}{\includegraphics{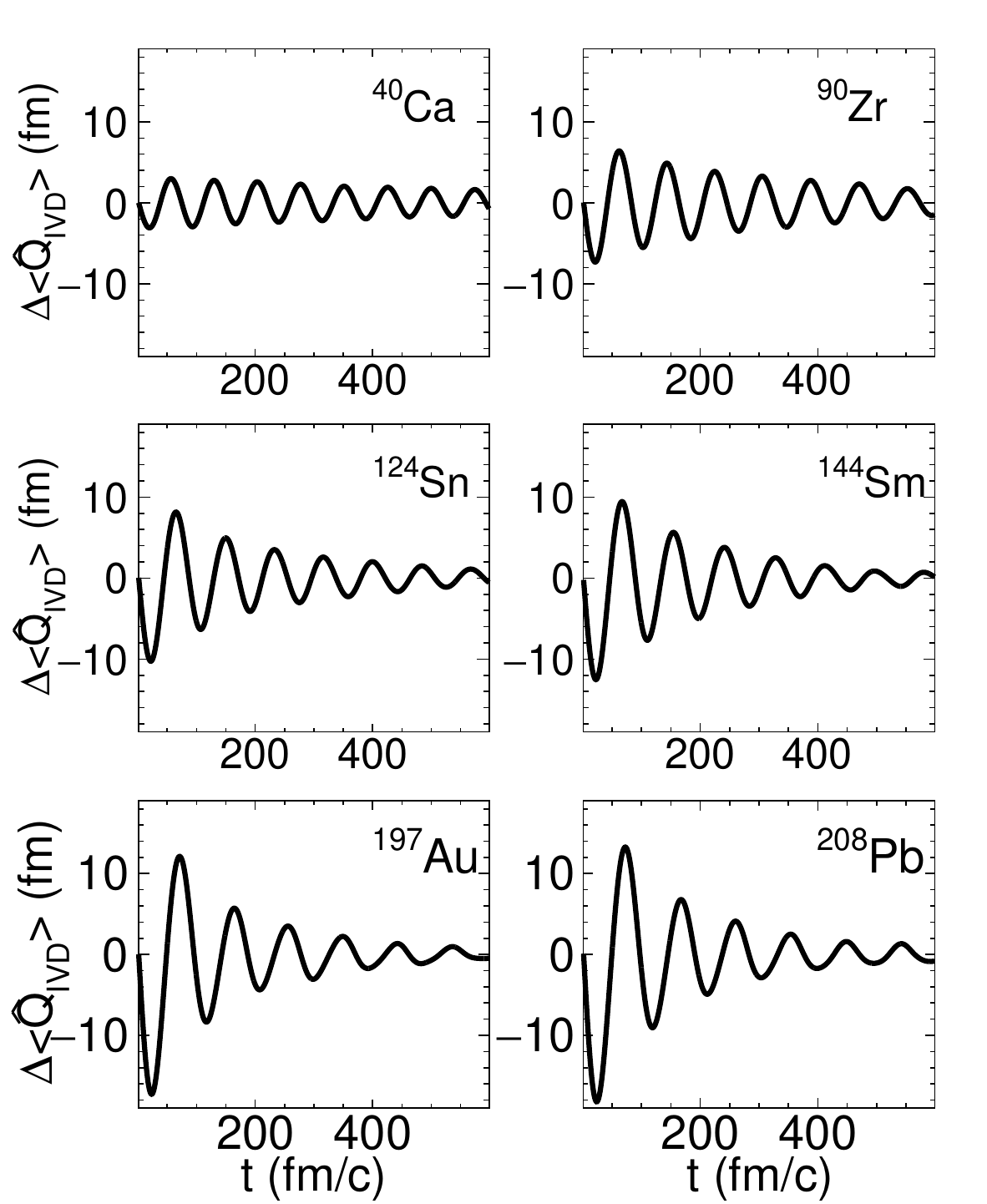}}
\caption{\label{fig:Ca to Pb} The GDR oscillations of $^{40}$Ca, $^{90}$Zr, $^{124}$Sn, $^{144}$Sm, $^{197}$Au and $^{208}$Pb. The  SLy6 parameter setting is used.}
\end{figure} 

\begin{figure}[htbp]
\resizebox{8.6cm}{!}{\includegraphics{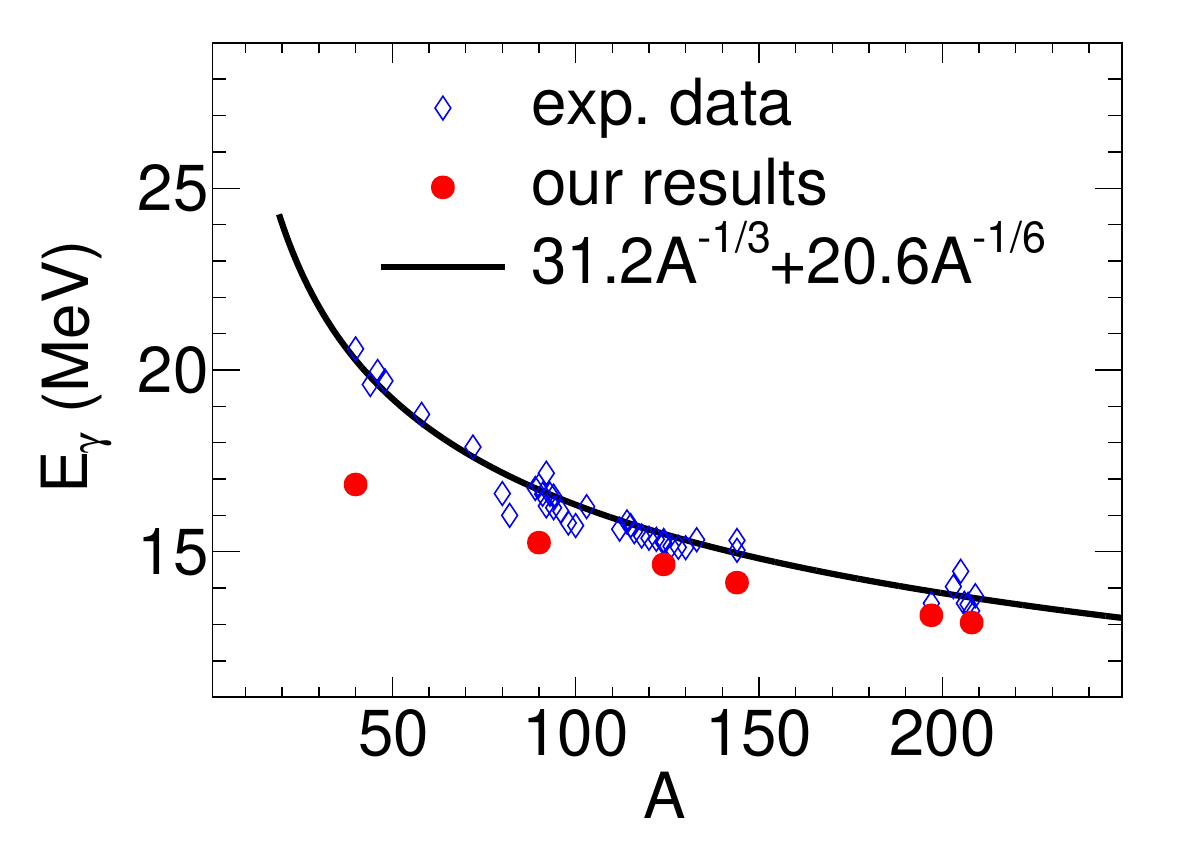}}
\caption{\label{fig:mass} GDR peaks plot as a function of the mass number. The red circles are our EQMD results utilizing the SLy6 parameter setting.  }
\end{figure} 

We also study the mass number dependence of GDR peak energy.
In Fig.~ \ref{fig:Ca to Pb}, the GDR oscillations of $^{40}$Ca, $^{90}$Zr, $^{124}$Sn, $^{144}$Sm, $^{197}$Au and $^{208}$Pb are presented. As expected, clear GDR signals are obtained in the framework of our EQMD.
The corresponding peak energy as a function of mass number is plotted in Fig.~\ref{fig:mass}.
The red circles are our results utilizing the SLy6 parameter setting. 
The blue diamonds are the experimental data \cite{exp_data}, and the black solid line is the well-known phenomenological formula, i.e., \cite{gamma} 
\begin{equation}
31.2A^{-1/3} + 20.6A^{-1/6}.
\end{equation}   
Compared with experimental data, our results underestimate the peak energy overall with the SLy6 parameter setting.
In the region of large mass numbers the deviation is relatively small, but becomes more pronounced in the region of small mass numbers. It should be noted that the parameters used in this work are not deliberately matched to the experimental results, and the effective interaction used is not a standard Skyrme energy density function.
On the other hand, this is a result of the competition between the mean field potential and the other effects, such as shell effect, pair effect, nuclear deformation, etc. \cite{taochen}.
The EQMD model, as a type of transport model, describes nucleons as independent particles moving in the mean field potential.
In the region of heavy nuclei, the mean field potential is dominant, so the collective oscillation of heavy nuclei can be reasonably described by the transport model.
In the region of light nuclei, on the other hand, other effects begin to emerge.
How to build a self-consistent description of the nuclear collective vibration from light to heavy nuclei is a difficult task for the transport model and beyond the scope of this article. 
Finally, we emphasize again the purpose of this article, which is to introduce a new numerical method for using a non-integer exponent of density integration variable in the EQMD model without causing the system to diverge or initialization failure.

\section{Conclusion\label{conclusion}}

The EQMD model is a powerful tool to study the effects of $\alpha$ clusters in heavy ion collisions.
However, in the original EQMD model, the variable $\gamma$ in the Skyrme potential can only take an integer value of 2, leading to inappropriate nuclear incompressibility.
Also, the form of the mean field potential is slightly simplified.
The propagation of the nucleons cannot be solved with a non-integer exponent for the integral of the density using a common approximation method, which always leads to initialization failure and system divergence in EQMD. 
In this article, a Monte Carlo integration algorithm is used to overcome this problem.
Then we use GDR as a tool, which requires that the nuclear system has high stability, to verify the correctness and advantage of the Monte Carlo integral for nucleon propagation.
Based on this advantage, we investigate the mass dependence of GDR, and the results are generally consistent with some existing experimental data.
Overall, we have successfully overcome the obstacles by introducing soft potential terms in EQMD, which is a crucial step for this model to explore more valuable effects for heavy ion collisions in the near future.

\section*{Acknowledgements}
This work is partially supported by the National Natural Science Foundation of China under Contracts No. $12347149$, $11890714$, and $12147101$，$11925502$, the Strategic Priority Research Program of CAS under Grant No. XDB34000000, the National Key R\&D Program of China under Grant No. 2022YFA1602300, and the Guangdong Major Project of Basic and Applied Basic Research No. 2020B0301030008.


\end{CJK*}
\bibliography{eqmd_and_gdr}

\end{document}